\documentclass[journal]{IEEEtran}     
\usepackage[table,xcdraw]{xcolor}
\usepackage{nomencl}
\makenomenclature
\usepackage{amssymb}
\usepackage{arydshln}
\usepackage{indentfirst}
\usepackage{filecontents}
\usepackage{lipsum}
\usepackage{tabularx}
\usepackage{array}
\usepackage[noadjust]{cite}
\usepackage{subfigure}
\usepackage{epsfig}
\usepackage{physics}
\usepackage{balance}
\usepackage{hhline}
\usepackage[normalem]{ulem}
\usepackage{amssymb,amsmath,color,graphicx, amsbsy}
\usepackage{algorithm}
\usepackage{algpseudocode}
\usepackage{amsfonts}
\usepackage{epstopdf}
\usepackage{multirow}
\usepackage{verbatim}
\usepackage{mathtools}
\usepackage{nomencl}
\usepackage{graphicx}
\graphicspath{{Figures_inr/}}
\usepackage{xcolor}
 \usepackage{soul}
\usepackage{lipsum,multicol}
\usepackage{multicol}
\usepackage{dblfloatfix}
\usepackage{float}
\usepackage{cuted}
\usepackage{multirow}
\edef\oldtt{\ttdefault}
\usepackage[scaled]{beramono}
\usepackage{courier}
\usepackage[T1]{fontenc}
\renewcommand{\ttdefault}{\oldtt}

\usepackage{etoolbox}
\usepackage{algpseudocode}

\usepackage{multirow}
\usepackage{subfigure}
\usepackage{enumitem}

\begin{document}

\IEEEpeerreviewmaketitle

\title{\fontsize{22pt}{27pt}\selectfont{Implicit Neural Representation of Waveform Measurements in Power Systems Waveform Data Analysis}}

\author{Narges Ehsani,~\IEEEmembership{Student Member, IEEE}, 
Vishwanath Saragadam, and Hamed Mohsenian-Rad,~\IEEEmembership{Fellow, IEEE}
\vspace{-0.3cm}
\thanks{The authors are with the  University of California, Riverside, CA, USA. 
The corresponding author is H. Mohsenian-Rad. E-mail: hamed@ece.ucr.edu.}\vspace{-0.3cm}
}

\maketitle

\begin{abstract}

There is currently a paradigm shift in several power system monitoring applications, such as incipient fault detection and monitoring inverter-based resources, to transition from traditional phasor analytics to more informative waveform  analytics.
This paper  contributes to this transition by developing a novel approach to modeling voltage and current waveform measurements using implicit neural representations (INRs). INRs are continuous function approximators that are recently used in vision and signal processing. The proposed INR models are specifically designed to meet the requirements of waveform analytics in power systems, such as by using sinusoidal activation functions that capture the  periodic nature of voltage and current waveforms. We also propose extended models that can efficiently represent correlated waveforms, such as  three-phase waveforms and synchro-waveforms. Real-world case studies demonstrate the effectiveness of the proposed INR models in terms of accuracy (<1-2\% MSE) and model size (4-6$\times$ compression). We also investigate the application of INR models in  oscillation monitoring, for   single mode oscillations and dual mode modulated oscillations.

   \vspace{0.1cm}

\end{abstract}

\vspace{-.1 cm}

\section{Introduction}

\subsection{Waveform Analytics: A New Big Data Challenge}

Waveforms are the most authentic and granular representation of voltage and current in power systems. Nevertheless, the traditional approach in power system monitoring has been to measure RMS values, phasors, or other filtered/processed representations of voltage and current; which results in losing some important details that are only visible in waveforms. 

However, with the recent advancements in power system
sensor technologies, including the advent of synchro-waveforms, and also because of the increasing complexity in power system operation caused by the high-penetration of inverter-based resources (IBRs), there is now a growing interest in conducting analysis and inference directly on waveform measurements; e.g., see the detailed discussions in \cite{NASPI_Report, hamed_magazine_2023, Wilsun_Xu_2022}. 

Waveform data are obtained from waveform measurement units (WMUs), through event-triggered waveform capture or continuous  streaming of waveform measurements \cite[Ch 4]{Hamed_book}. The latter is similar to how phasor measurements are streamed by phasor measurement units (PMUs). However, WMUs report data at a much higher rate than PMUs. For example, each three-phase WMU reports 3,981,312,000 voltage readings per day (at a sampling rate of 256 samples per cycle), which can exceed one gigabyte of data per day per sensor \cite{hamed_magazine_2023, Narges_NAPS}.

Collecting data at such high reporting rate creates a new challenge in
big data analytics in power systems, moving beyond the existing big data analytics practice for traditional measurements such as from smart meters and PMUs.

\subsection{Approach and Contributions}

We seek to address some of these challenges by developing a novel approach to model waveform (and synchro-waveform) measurements. We propose to utilize the latest advancements in the field of implicit neural representation (INR). 
INRs are continuous function approximators based on multilayer perceptrons (MLPs).
Since their first widespread usage in novel view synthesis in graphics~\cite{mildenhall2020nerf}, INRs have been quickly adopted in all fields of vision and signal processing,  including rendering~\cite{kuznetsov2021neumip}, medical imaging~\cite{wang2022neural}, and virtual reality~\cite{deng2022fov}.

\vspace{0.1cm}

The main contributions in this paper are as follows:

\begin{itemize}[leftmargin=*]
    \item To the best of our knowledge, this is the first study to develop INR models for waveforms in power systems. 
    The proposed INR models are designed to meet the requirements in this new application domain, such as by using sinusoidal activation functions~\cite{sitzmann2020implicit} that can capture the inherently periodic nature of voltage and current waveforms.

\vspace{0.05cm}

    \item The INR models with single and double hidden layers are analyzed. {We show that a single hidden layer INR resembles a Fourier transform and hence cannot capture transient distortions. In contrast, we show that a double hidden layer INR does capture these complexities. This enables an approximately $3\times$ increase in accuracy for the same number of parameters as a single hidden layer INR}. 

\vspace{0.05cm}

    \item \textcolor{black}{Unlike classical signal representations, INRs enable representing correlated signals with a single model}. We hence extend our INR  model to enhance efficiency in modeling correlated waveforms, such as among three phases, among synchro-waveforms, and between voltage and current. \textcolor{black}{Experiments on real-data demonstrate that such an approach is highly advantageous in modeling sub-cycle transients that are otherwise difficult to model with separate INRs.}

\vspace{0.05cm}

    \item Real-world case studies provide detailed sensitivity analysis and confirm the performance of the proposed INR models, both in terms of model accuracy and model size. 
    The direct application of the INR models is investigated in analyzing two types of oscillatory events, namely single mode oscillations and dual mode modulated oscillations.
    
\end{itemize}

\subsection{Related Literature}

In power systems, research on INRs is still emerging, with limited studies exploring its potential. In \cite{acceleration_2024}, INRs are used for power system dynamic simulations. In \cite{network_flow_2022, optimal_PF_2021}, INRs are used in optimal power flow analysis. In \cite{load_forcast}, INRs are utilized for load forecasting. While these studies demonstrate INR’s potential in power systems, they highly differ from the focus of our research. Here, we leverage INRs not for optimization or solving partial differential equations, but to compactly represent high-resolution power system waveforms.

The analysis in this paper also falls under the growing literature on waveform data analytics in power systems. The advent of synchro-waveform technology has given a recent boost to this field, with  applications, such as for fault location identification \cite{Imtiaj_Khan_2023}, wildfire monitoring \cite{Hamed_PQ_SDGE},  oscillation monitoring \cite{Taimur_Zaman}, and dynamic modeling  of IBRs \cite{fatemeh_letter_journal}.

The scope of this paper also has some partial overlap with the literature on data compression, such as in \cite{slipstream_2022} for loss-less compression, often with focus on generic data compression techniques, or in \cite{subband_2021} for lossy compression, often with focus on signal processing, such as Fourier and Wavelet transform. Although our proposed INR models can significantly compress the waveform data, they are not designed for data compression, but rather to provide a systematic way to model waveforms.

\section{INR Modeling of Waveform Measurements with Single or Double Hidden Layers}\label{sec:INR_I}

\begin{figure}[t]
    \centering
    \vspace{-0.2cm}
    \includegraphics[width=0.85\columnwidth]{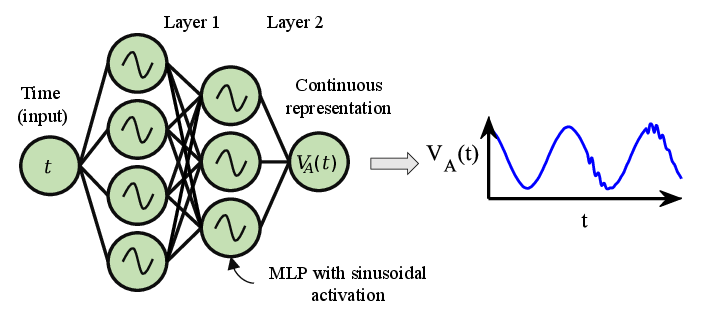}
    \vspace{-0.1cm}
    \caption{The  proposed INR architecture to model a single time-series of waveform measurements. The MLP has two hidden layers. The activation functions at all neurons in both hidden layers are sinusoidal functions.}
    \label{fig:siso_diagram}
    \vspace{-0.2cm}
\end{figure}

The overall architecture of the first proposed INR model is shown in Fig. \ref{fig:siso_diagram}. The input to the model is time $t$. The output is waveform $x(t)$, such as the time-series of voltage waveform measurements at Phase A,  denoted by $v_A(t)$. The nonlinear activation function $\sigma(\cdot)$ plays a key role in the representation capacity of the INR model. In this paper, given the dominantly sinusoidal nature of the waveform measurements in power systems, we use $\sigma(\cdot) = \sin(\cdot)$. Sinusoidal activation functions are proven to provide significantly higher representation accuracy than ReLU, in particular for signals with periodic nature \cite{sitzmann2020implicit}. 

\subsection{Single Hidden Layer INR}

Suppose the INR model has only one hidden layer. Let $\mathbf{a}_1, \mathbf{b}_1$ and $\mathbf{a}_2, \mathbf{b}_2$ be the weights of the first layer and the second layer, respectively. We can obtain the output signal as 
\begin{equation}
    x(t) = \sum_{i=1}^h a_{2,i} \sin\left( a_{1, i}t + b_{1,i}\right) + b_2, 
    \label{eq:siso}
\end{equation}
where $h$ denotes the number of neurons in the hidden layer. 
The number of parameters in the INR model in (\ref{eq:siso}) is:
\begin{equation}
    2h + h + 1 = 3h + 1.
\end{equation}
The formulation in (\ref{eq:siso}) can be interpreted as a Fourier transform, where for each $i = 1, \ldots, h$, parameter $a_{1,i}$ acts as  harmonic frequency, $a_{2,i}$ acts as harmonic magnitude, and   $b_{1,i}$ acts as  harmonic phase angle. Therefore, the INR model with one hidden layer is comparable with the \emph{phasor representation} of the waveform measurements. However, unlike phasors, which focus primarily on the fundamental frequency of the power system, the model in (\ref{eq:siso}) is \textit{not} biased on any particular frequency. All the parameters are rather \emph{trained} by a stochastic gradient descent approach. Despite this advantage, an INR with a single hidden layer cannot significantly improve the model accuracy compared to Fourier transform; because the formulation in (\ref{eq:siso}) is still very similar to Fourier representation. 

\vspace{-0.1cm}

\subsection{Double Hidden Layer INR}

Next, suppose the INR model has two hidden layers. This is the same setup as in Fig. \ref{fig:siso_diagram}. Suppose the number of neurons in the first hidden layer and in the second hidden layer are $h_1$ and $h_2$, respectively. Let $\mathbf{a}_1, \mathbf{b}_1$ and $\mathbf{a}_2, \mathbf{b}_2$ and $\mathbf{a}_3, \mathbf{b}_3$ be the weights of the first layer and the second layer and the third layer, respectively.
We can obtain the output signal as 
\begin{equation}
    x(t) = \sum_{j=1}^{h_2}
a_{3,j} \sin \left(
\sum_{i=1}^{h_1} a_{2,i, j} \sin\left( a_{1, i}t + b_{1,i}\right) + b_{2, i} \! \right) \! + \! b_3. 
  \label{eq:simo}
\end{equation}

The model in (\ref{eq:simo}) does no longer resemble Fourier transform. Intuitively, the second hidden layer introduces higher spectral diversity, which enables a richer representation in (\ref{eq:simo}) than in (\ref{eq:siso}). Indeed, as explained in \cite{roddenberry2023implicit}, INRs with two or more hidden layers provide a larger convergence of the frequency space than INRs with only one hidden layer. Since voltage and current waveform measurements in power systems have several higher harmonics and transient modes, there is great potential to leverage INRs based on the architecture in Fig. \ref{fig:siso_diagram}, to represent waveform measurements in power systems. 

The number of parameters in the INR model in (\ref{eq:simo}) is:
\begin{equation}
    2 h_1 + h_1 h_2 + 2 h_2 + 1. 
\label{parameters:siso}
\end{equation}
As we will see next, this increase in parameter count dramatically increases representation accuracy of the waveforms.

\subsection{Initial Case Study}

\begin{figure}[t]
    \centering
    \vspace{-0.2cm}
\includegraphics[width=0.47\textwidth]{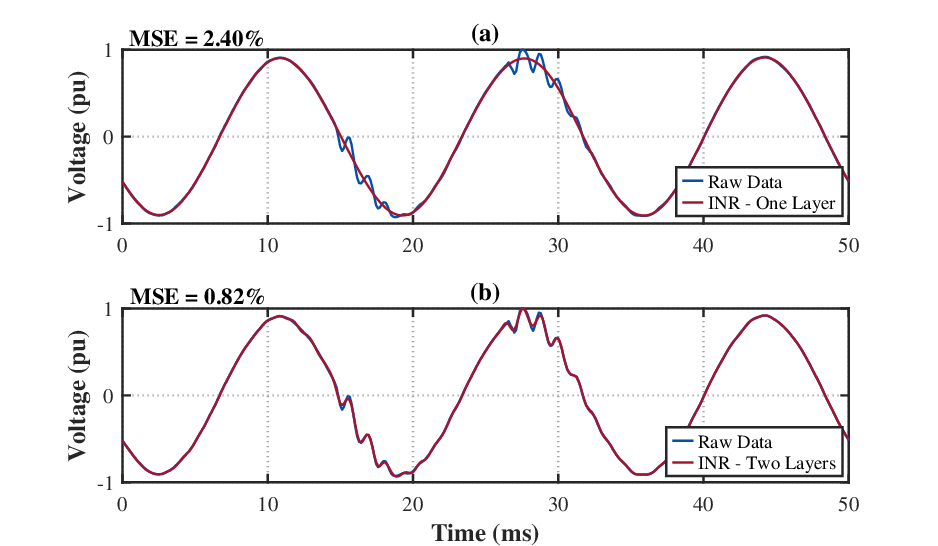}
    \vspace{-0.3cm}
    \caption{Model accuracy in reconstructing a real-world voltage waveform measurement with a sub-cycle oscillatory event: (a) using the INR model with single hidden layer; (b) using the INR model with double hidden layers. The two INR models in this example have equal number of parameters.}
    \label{fig:single_vs_double}
    \vspace{-0.3cm}
\end{figure}

Fig. \ref{fig:single_vs_double} provides an example to derive INR representations for a real-world voltage waveform measurement that contains a sub-cycle oscillatory event. The raw waveform measurements capture 62 cycles, with two event cycles in the middle. Here, we only show the portion of the waveform that contains event. 

In Fig. \ref{fig:single_vs_double}(a), the INR model has one hidden layer. The number of  parameters is $1663 = 3 \times 554 + 1$, where  $h = 554$.  The Mean Squared Error (MSE) is 2.40\%, which is calculated between the raw waveform measurement (blue) and the reconstructed waveform (red) using the model in (\ref{eq:siso}). The MSE is calculated across the entire 62 cycles of the waveform measurements. Although this model can effectively approximate the steady-state signal,  resembling the Fourier series, it lacks the capability to capture the more complex behavior in the transient component of the signal. 

In Fig. \ref{fig:single_vs_double}(b), the INR model has two hidden layers, with $1661 = 
2 \times 30 + 30 \times 50 + 2 \times 50 + 1$ parameters, where $h_1 = 30$ and $h_2 = 50$. Thus, the number of parameters in the two-layer INR in Fig. \ref{fig:single_vs_double}(b) is equal to the number of parameters in the one-layer INR in Fig. \ref{fig:single_vs_double}(a). 
However, the MSE in the two-layer model is 0.82\%, which is drastically less than the MSE of 2.40\% for the equal-sized INR model with one hidden layer.

We also evaluated the impact of using ReLU as an alternative to sine activation. The results showed that an INR model of the same size and architecture but with ReLU activation performed poorly, with an MSE of 18.88\%. Accuracy was particularly low during steady-state behavior, as ReLU struggles to capture the waveform’s periodic nature. Performance was also poor during transients, with an MSE of 1.56\% for ReLU compared to only 0.48\% for sine activation. Thus, the sinusoidal activation function clearly outperforms ReLU. 

\section{Extended Model to Enhance Efficiency in Three-Phase Waveforms and Synchro-Waveforms}\label{sec:INR_II}

\begin{figure}[t]
    \centering
    \vspace{-0.2cm}
    \includegraphics[width=\columnwidth]{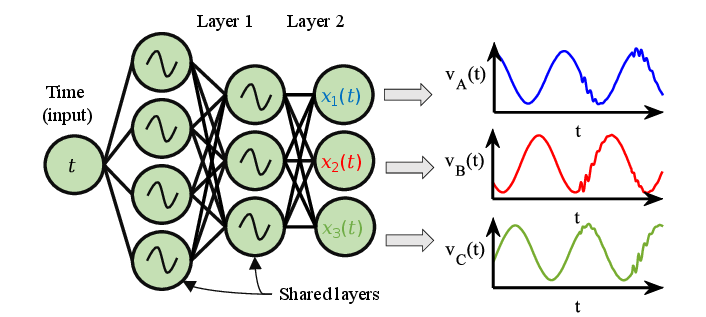}
    \vspace{-0.5cm}
    \caption{The proposed model to enhance efficiency in modeling three-phase waveform measurements. A similar extension can help improve model efficiency when it comes to synchronized waveform measurements.}
    \label{fig:simo_diagram}
    \vspace{-0.1cm}
\end{figure}

The waveform measurements in power systems are often correlated. Correlation could exist among different waveforms, including voltages across three phases, voltages across multiple locations, and voltage and current waveforms at a given location. %:
Such correlations may help reduce the size of the INR models. For instance, suppose we seek to model voltage waveform measurements on Phase A, Phase B, and Phase C. The waveform measurements are denoted by $v_A(t)$, $v_B(t)$, and $v_C(t)$, respectively. We have two options to derive the models:
\begin{itemize}
    \item Develop \emph{three separate} INR models based on Fig. \ref{fig:siso_diagram}. 

    \item Develop \emph{one combined} INR model based on Fig. \ref{fig:simo_diagram}.
    
\end{itemize}

The first option simply repeats the INR architecture in Fig. \ref{fig:siso_diagram} three times, \emph{once for each phase}. The number of parameters  would be three times the number of parameters for one phase.
\begin{equation}
    3\times(2h_1 + h_1 h_2 + 2h_2 + 1)
\label{parameters:siso:three}
\end{equation}

In the second option, we use \textit{one} INR model with \textit{three} outputs, one for each phase. In this new architecture, the two hidden layers are \emph{shared} between the three outputs; as shown in Fig. \ref{fig:simo_diagram}.
The number of parameters in this option is:
\begin{equation}
    2h_1 + h_1 h_2 + h_2 + 3 \times (h_2 + 1)\color{black}. 
\label{parameters:simo:three}
\end{equation}
\noindent This results in significantly fewer parameters than in (\ref{parameters:siso:three}).

We can similarly apply the above approach to other cases with correlated measurements, including synchro-waveforms. 
For example, suppose we need to model the \emph{time-synchronized} voltage measurements on the same phase but at five different locations. 
One option is to develop \emph{five separate} INR models based on Fig. \ref{fig:siso_diagram}. The other option is to develop \emph{one combined} INR model based on the architecture in Fig. \ref{fig:simo_diagram}, with five (instead of three) neurons in its output layer.

\begin{figure*}[t]
    \centering
    \vspace{-0.2cm}
    \hspace*{-1.5cm}
    \includegraphics[width=0.96\textwidth]{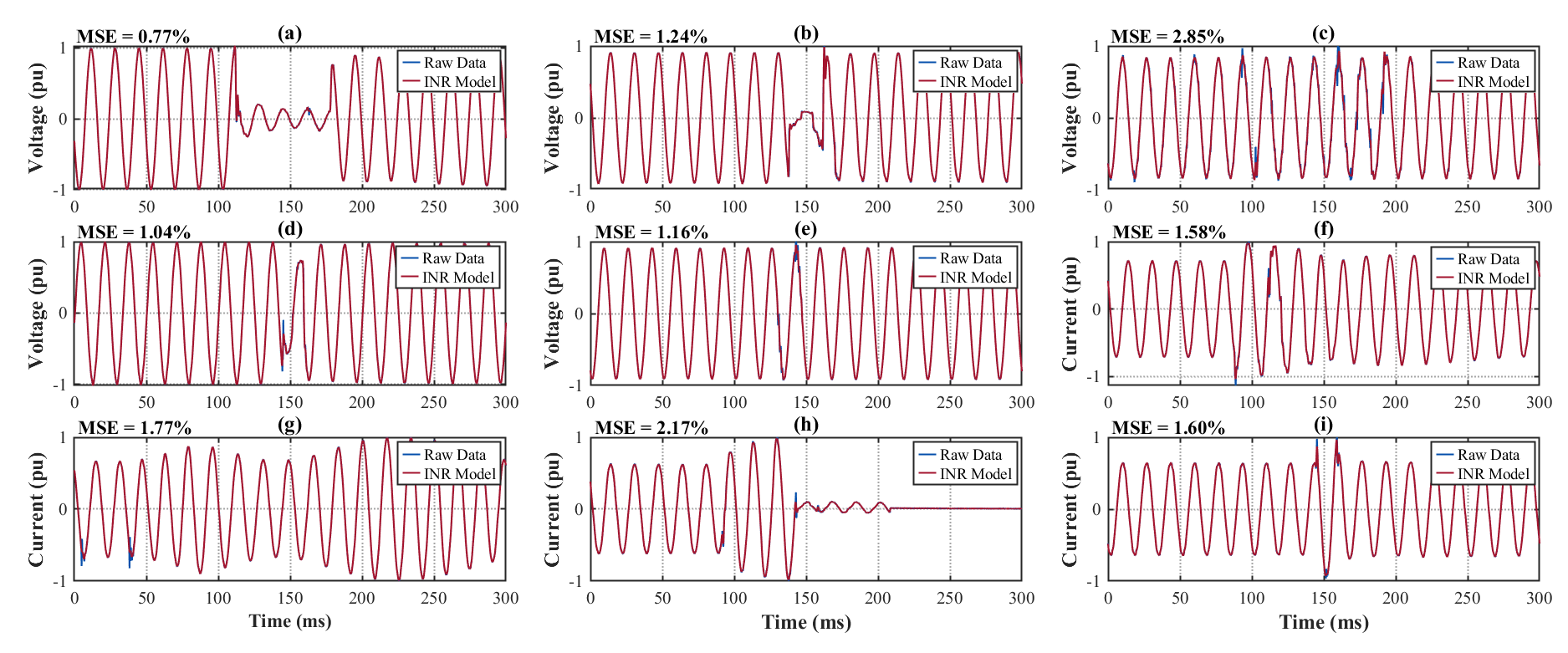}
    \hspace*{-1.5cm}
    \vspace{-0.4cm}    
    \caption{\color{black} Various waveform event signatures. Each example shows the raw waveform measurements, the corresponding reconstructed waveforms from an INR model with double hidden layers, and the corresponding MSE: (a)-(e) event signatures in voltage waveforms; (f)-(i) event signatures in current waveforms.}
    \vspace{-0.3cm}
    \label{fig:examples}
\end{figure*}
\color{black}

\section{Additional Real-World Case Studies}\label{sec:applications}

\vspace{0.1cm}

\subsection{Modeling Diverse Waveform Signatures}

Fig. \ref{fig:examples} shows examples of real-world waveform measurements, including raw data (blue) and reconstructed waveforms (red) based on INR with $h_1 = h_2 = 50$. The measurements are obtained from a three-phase SEL 735 power quality sensor at 480V (line-to-line). The sampling rate of the waveform measurements is 128 samples per cycle, i.e., 128 × 60 = 7680 samples per second. Each waveform capture includes 62 cycles of waveform measurements, centered around the start of the event. The event signatures are diverse. They include both voltage waveforms, in Figs. \ref{fig:examples}(a)-(e), and current waveforms, in Figs. \ref{fig:examples}(f)-(i).
The MSE varies from 0.77\% to 2.85\%.

\subsection{Sensitivity Analysis: Number of INR Parameters}

\begin{figure}[t]
    \centering
    \vspace{-0.1cm}
\includegraphics[width=0.47\textwidth]{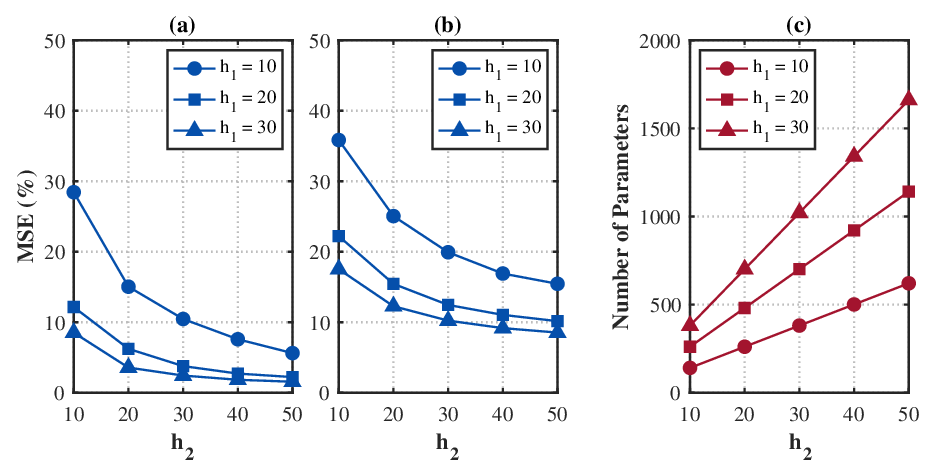}
    \vspace{-0.2cm}
    \caption{ Sensitivity analysis of the INR models to the number of neurons in each layer, namely $h_1$ and $h_2$: (a) The MSE in modeling voltage waveforms; (b) The MSE in modeling current waveforms; (c) The size of the model.}
    \label{fig:sensitivity}
    \vspace{-0.3cm}
\end{figure}

Fig. \ref{fig:sensitivity} provides a detailed sensitivity analysis with regards to the number of neurons in the first and second hidden layers, denoted by $h_1$ and $h_2$, respectively. 
The results in Fig. \ref{fig:sensitivity}(a) and (b) provide the MSE in modeling \emph{voltage} and \emph{current} waveforms, respectively. The results in Fig. \ref{fig:sensitivity}(c) provide the corresponding model sizes, which are the same for both voltage and current waveform for the same choices of $h_1$ and $h_2$. 
Each point in any curve in Figs. \ref{fig:sensitivity}(a) and (b) is the average of $30 \times 10 = 300$ values, from 30 distinct waveform signatures and 10 runs of INR model training for each event signature.  
We can make several important observations based on the results in Fig. \ref{fig:sensitivity}. First, there is a clear trade-off between  the accuracy and the size of the  model. Increasing both $h_1$ and $h_2$ can improve model accuracy but increase the model size. Second, increasing $h_2$ appears to help more than increasing $h_1$ in order to improve model accuracy. Third, whether we increase  $h_1$ or $h_2$, the gain in model accuracy demonstrates gradual saturation. Fourth, the model accuracy is clearly higher for voltage waveforms than current waveforms. This is because, in practice, current waveforms are often more volatile and distorted than voltage waveforms, as we  saw in Fig. \ref{fig:examples}.  

The computation time for training an INR model depends on the size of the model. The smallest INR model with $h_1 = 10$ and $h_2 = 10$ takes 3.63 seconds to train. The largest INR model with $h_1 = 50$ and $h_2 = 70$ takes 7.70 seconds to train. Training is done on Google Cloud Platform (GCP) using a Tesla T4 GPU with 15 GB of memory and 28 GB RAM. 

\subsection{Three Separate versus One Combined INR Model}

\begin{figure}[t]
    \centering
    \vspace{-0.2cm}
\includegraphics[width=0.47\textwidth]{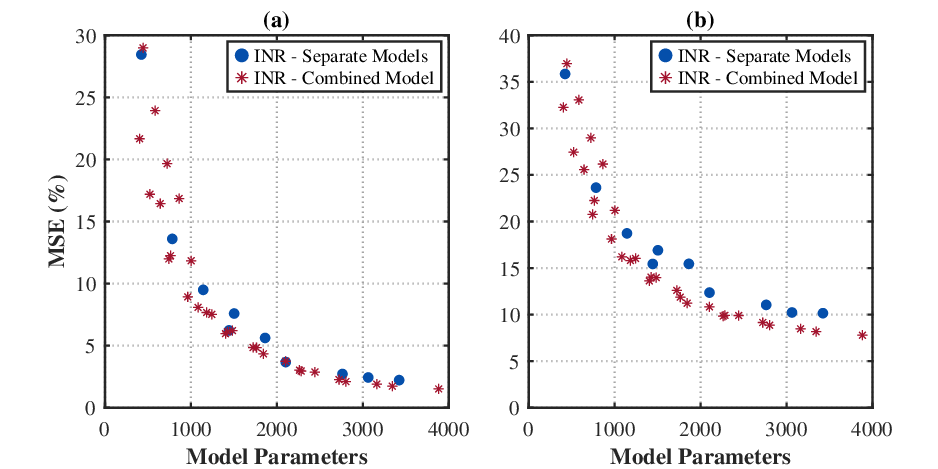}
    \vspace{-0.1cm}
    \caption{ Comparing the performance of three separate INR models in Fig. \ref{fig:siso_diagram} versus one combined INR model in Fig. \ref{fig:simo_diagram} in modeling three-phase waveform measurements. The MSE for each model is averaged over three phases and across several runs: (a) Average MSE for voltage waveforms across three phases; (b) Average MSE for current waveforms across three phases.}
    \label{fig:single_multiple}
    \vspace{-0.3cm}
\end{figure}

Recall  that we have two options to model three-phase waveform measurements: using three separate models as in Fig. \ref{fig:siso_diagram} or one combined model as in Fig. \ref{fig:simo_diagram}. Fig. \ref{fig:single_multiple} provides a comparison between these two options. The MSE is obtained for each approach for a varying number of parameters. Each point represents the average of $3 \times 30 \times 10 = 900$ experiments, calculated from three phases of 30 distinct waveform signatures, and 10 runs of INR model training for each event. The blue points correspond to the INR model in Fig. \ref{fig:siso_diagram}, where the average output of three separate models is used. The red points correspond to the combined INR model in Fig. \ref{fig:simo_diagram}. 

The results in Fig. \ref{fig:single_multiple}(a) are for the INR models of the voltage waveform measurements, and the results in Fig. \ref{fig:single_multiple}(b) are for the INR models of the current waveform measurements. 

\vspace{-0.1cm}

\subsection{Application in Analysis of Waveform Oscillations}

Lastly, we use the INR models in the analysis of real-world 

\noindent waveform oscillations. We study two types of oscillatory 
behavior, that have fundamentally different characteristics: 
\begin{itemize}[leftmargin=*]
    \item Single mode oscillations, where the frequency of the dominant mode of oscillation is denoted by $f_\text{dominant}$.

    \vspace{0.1cm}
    
    \item Dual mode \emph{modulated}  oscillations, where the oscillations occur at a \emph{pair} of \emph{sideband} frequencies $60 \text{ Hz} \pm f_\text{sideband}$. 
\end{itemize}

\begin{figure}[t]
    \centering
    \vspace{-0.2cm}
\includegraphics[width=\columnwidth]{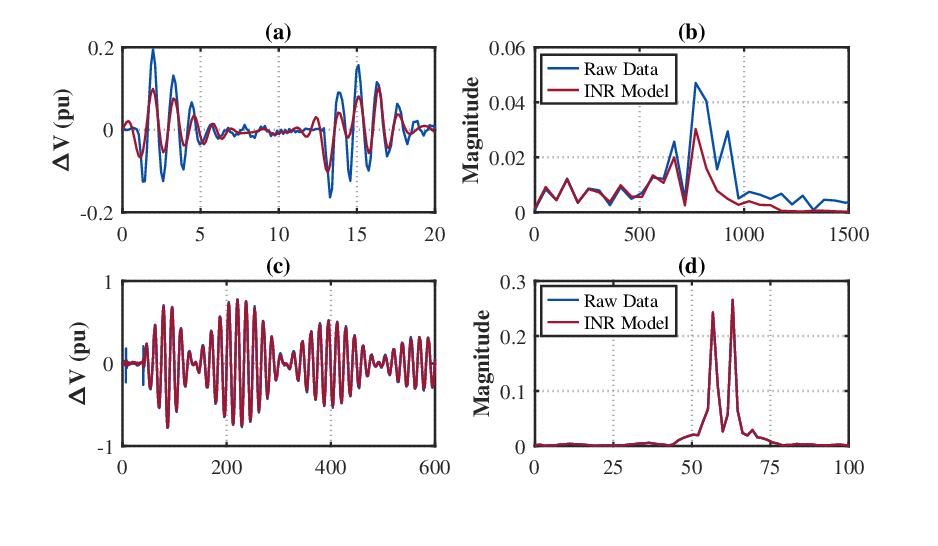}%
\vspace{-0.7cm}
\includegraphics[width=\columnwidth]{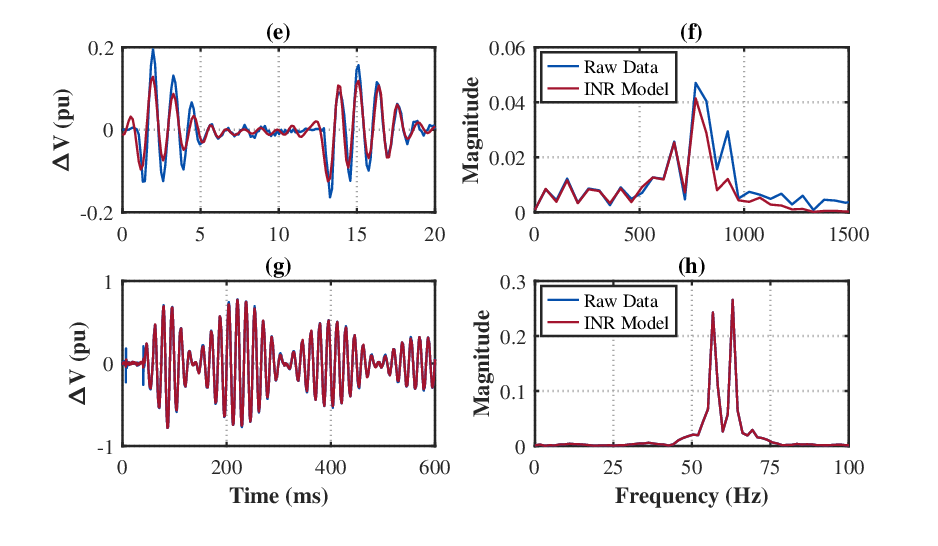}
    \vspace{-0.8cm}
    \caption{
    Analysis of the frequency spectrum (in the right column) for two types of oscillations and two types of methods: (a)(b) single mode oscillation based on separate INRs; (c)(d) dual mode modulated oscillation based on separate INRs; (e)(f) single mode oscillation based on a combined INR; and (g)(h) dual mode modulated oscillation based on a combined INR.}
    \label{fig:FFT_analysis}
    \vspace{-0.2cm}
\end{figure}

Both analysis require examining the \emph{frequency spectrum} of the waveforms. Thus, we compare the frequency spectrum from the raw data, as well as the INR output, using Discrete Fourier Transform (DFT).  %:
The analysis is done on all phases. The results on Phase A are shown   
in Fig. \ref{fig:FFT_analysis}. 
The raw data are \emph{differential waveforms}, which are the extracted superimposed signature of the event on the normal waveform \cite[Section 4.2.5]{Hamed_book}. Each row shows the waveforms in time domain on the left side, and the derived frequency spectrum on the right side.  

The results in Figs. \ref{fig:FFT_analysis}(a)-(d) for the INR model in Fig. \ref{fig:siso_diagram}, where three separate INR models are obtained for the three phases. The number of parameters in the INR model across all phases is 8103. 
The results in Figs. \ref{fig:FFT_analysis}(e)-(h) for the INR model in Fig. \ref{fig:simo_diagram}, where a combined INR model is obtained for all three phases. The number of parameters in the INR model for all phases is 5503. 
The total number of parameters in the raw data for all phases in each case study is 23,808, thus achieving $4-6\times$ compression with INR models. 

Fig. \ref{fig:FFT_analysis}(a, b, e, f) show the time and frequency content of a single-mode oscillatory event. Both separate and combined INRs accurately capture the dominant frequency $f_\text{dominant}$ at 900 Hz, consistent with the raw data. Importantly, a combined INR represents the spectrum more accurately, with fewer parameters, 
enabling a more parameter-efficient representation.

Figs. \ref{fig:FFT_analysis}(c, d, g, h) visualize the results for dual mode modulated oscillatory event. Both separate and combined INRs accurately model the spectra, with clear peaks at the two side band frequencies $\pm f_\text{sideband}$ around 60 Hz.
Importantly, the combined INR enables higher accuracy with fewer parameters. 

\section{Conclusions and Future Work}\label{sec:conclusion}

INR is shown to be a powerful method to model voltage and current waveform measurements in power systems. A single hidden layer INR
resembles a Fourier transform and hence cannot capture
transient distortions. In contrast, a double
hidden layer INR provides a novel foundation to capture these complexities in power systems waveforms. In most scenarios, this
enables an approximately $3\times$ increase in accuracy for the
same number of parameters as a single hidden layer INR.

The proposed INR models are modified to model correlated waveform measurements with significantly fewer parameters, such as in simultaneously  modeling voltage on all three phases. 

Detailed sensitivity analysis was conducted based on real-world data to identify the importance of parameters, as well as to demonstrate the performance of the proposed INR models in working with both voltage and current waveform measurements during various events in power systems.

Importantly, the proposed INR models  demonstrated high accuracy in modeling the key waveform characteristics not only in time domain but also in frequency domain, providing a direct application in any event characterization tasks. 

This study can be extended in multiple directions. While our analysis considers up to two hidden layers in the INR architecture, exploring deeper networks and alternative architectures may further enhance model accuracy. The robustness of INR models could also be assessed more extensively under a wider range of operating conditions, including varying noise levels and diverse event signatures. A more comprehensive comparison with state-of-the-art waveform modeling techniques can further clarify the advantages and limitations of INR models. Future work may also focus on integrating INR models into power systems monitoring and operation use cases, to support both off-line and on-line monitoring applications.

\bibliographystyle{IEEEtran}
\bibliography{References}

\end{document}